\documentstyle[aps,epsfig]{revtex} 
\newcommand{\be}{\begin{equation}} 
\newcommand{\ee}{\end{equation}} 
\newcommand{\ben}{\begin{eqnarray}} 
\newcommand{\een}{\end{eqnarray}} 
\begin{document}

\twocolumn[\hsize\textwidth\columnwidth\hsize\csname 
@twocolumnfalse\endcsname 

\title{
\normalsize
\mbox{ }\hspace{\fill}
\begin{minipage}{5.0cm}
UPR-940-T\\
{\tt hep-th/0105296}{\hfill}
\end{minipage}\\[5ex]
\centerline{\large\bf Network of Domain Walls on Soliton Stars}
}
\author{Francisco A. Brito$^{a}$ and Dionisio Bazeia$^b$} 
\address{$^a$Department of Physics and Astronomy\\
University of Pennsylvania, Philadelphia, Pennsylvania 19104, USA\\
$^b$Departamento de F\'\i sica, Universidade Federal da Para\'\i ba,\\ 
Caixa Postal 5008, 58051-970 Jo\~ao Pessoa, Para\'\i ba, Brazil}

\date{\today} 
 
\maketitle

\begin{abstract}
We explore the idea of a network of domain walls 
to appear at the surface of a soliton star. We show that for a suitable
fine tuning among the parameters of the model we can find localized fermion
zero modes only on the network of domain walls. 
In this scenario the soliton star gets unstable and decays into free
particles before the cold matter upper mass limit is achieved. However,
if fermions do not bind to the network of domain walls, the network becomes
neutral, imposing a new lower bound on the charge of the soliton star,
slightly raising its critical mass.
\end{abstract} 
\vskip2pc]

\newpage 

\section{Introduction}

According to general relativity, if a star has sufficiently
small mass, it can reach final equilibrium as white dwarf or 
a neutron star. On the other hand, if the mass of the collapsing
portion of a star is greater than the {\it cold matter} upper mass
limit $M_c$, the equilibrium can never be achieved and complete gravitational
collapse will occur. Usually, the final stage of such collapse is the formation
of a black role. For normal matter, the mass limit $M_c$ is approximately
equals to five solar masses $M_\odot$ (at zero angular
momentum) \cite{whee,wald}.

In the context of soliton stars it is possible to have objects
with much more large mass without gravitational collapse.
The subject of soliton stars has been introduced in Ref.\cite{tdl1} 
(See also \cite{tdl2,tdl3}). They are new types of cold stable stellar
configuration, and depending on the theory one may have
$M_c\sim 10^{15} M_\odot$. Recently \cite{dtcl} one has used
soliton star models in order to explain current observational
data, which favor the possible existence of a single supermassive
object at the center of our galaxy, other than a supermassive black hole.
 
In this paper we investigate how the structure of a soliton star can be
affected by the entrapment of another object on its surface. The idea that
we explore is similar to the case of lower dimensional domain walls living
inside domain walls \cite{mk,mr1,bsr,bb97,bb98,mr3,bbb}, which is inspired
by the mechanism used to build superconducting cosmic strings \cite{wit85}.
The basic mechanism consists in the formation of a scalar condensate due to
the spontaneous symmetry breaking of a scalar field, invariant under $U(1)$
gauge symmetry, confined to the cosmic string. In the context of brane
world scenarios renewed recently in \cite{nima,anton}
(see also \cite{akama,rubakov} for earlier ideas related to this topic),
one assumes that the fields describing the fundamental
particles, except the graviton (see Ref. \cite{rs2} for trapping of the
graviton; see also Ref. \cite{cv2k} and references therein for investigations
in supergravity), are confined to a 3-brane (a four-dimensional manifold
describing our universe) embedded in a higher dimensional spacetime. 
The scalar fields that live in the 3-brane can also generate a scalar
condensate by spontaneous symmetry breaking of some discrete symmetry.
When a discrete symmetry is spontaneously broken, for instance a $Z_2$
symmetry, a {\it domain wall} can form inside the 3-brane \cite{rgp}.
For other $Z_N$ symmetries, with $N>2$ we have the possibility of having
intersection of domain walls forming junctions, 
and then a network of domain walls. 

In order to investigate the entrapment of a network of domain walls by a
soliton star, we consider the possibility of a spherical \cite{oku,col}
two-dimensional wall to entrap wall segments that form a network. This
possibility may give rise to a network of domain walls \cite{bbr00a,bb00b}
to live at the surface of a standard soliton star \cite{tdl1}. We examine
this idea starting with an appropriate model, described by three real
scalar fields, introduced according to the lines of Ref.~{\cite{bb00b}}.
The model comprises several parameters, and below we show that depending
on the type of fine tuning used to adjust these parameters, we can produce
either heavier soliton stars or instability that will ultimately induce
their complete decay.

Our work is organized as follows. In Sec. \ref{mod} we present the model
and we investigate the entrapment of the network of domain walls.
In Sec. \ref{fnt} we study in detail the presence of localized zero modes
on the network and its consequences. Comments and conclusions are given
in Sec. \ref{conc}, which closes our work. Our notation is standard, and
we use dimensional units such that $\hbar\!=\!c\!=\!1$, and metric tensor
with signature $(+---)$. 

\section{The model}
\label{mod}

In Ref.~{\cite{bb00b}} we have investigated the possibility of a domain wall
to entrap a network of domain walls. This investigation was inspired in
Ref.~{\cite{bbr00a}}, which have dealt with the idea of building a planar
network of domain walls. The model there investigated engenders the $Z_3$
symmetry, which is the simplest symmetry that allows the presence of junctions
of domain walls. The presence of triple junctions in supersymmetric models
engendering the $Z_3$ symmetry were investigated
in Refs.~{\cite{gt,ct,sf,bb}}, with several distinct motivations. For
instance, the basic idea of Ref.~{\cite{gt}} was to present
Bogomol'nyi equation for the triple junction, showing that the planar
junction of domain walls only preserves $1/4$ supersymmetry of the model,
which is in contrast with the case of a single domain wall, which is known
to preserve $1/2$ supersymmetry of the corresponding model.

The presence of planar triple junctions may allow the tiling of the plane
with a regular hexagonal network, and this issue was further examined in
Ref.~{\cite{sf}}, and also in \cite{bbr00a,bb}. See
Refs.~{\cite{sh,bn,ino1,ag,na}}, for several other issues
related to this subject. As one knows, the most efficient
way to tile the plane with regular polygons is obtained by the
regular hexagonal network, and this brings the $Z_3$ symmetry as
the preferable symmetry, among many other possibilities. Very interestingly,
the $Z_3$ symmetry also appears as the center of the $SU(3)$ group, which
governs the symmetry of QCD, the field theory that describes strong
interactions; see, e.g., Ref.~{\cite{binosi}} and references therein
for recent investigations on this subject. The interest in walls and in
wall junctions widens when one recalls that the low energy world volume
dynamics of branes in string/M-theory may be described by standard models
in field theory \cite{m,w,gkp,s,gk}. Furthermore, this interest goes beyond
the context of high energy physics: For instance, it also appears in
ferroelectric materials where walls and wall junctions spring as stable
structures in many different ferroelectric crystals \cite{sal}.

The underlying symmetry of the standard model of elementary particles
contains the $SU(2)\times SU(3)$ group, and this is the basic inspiration
to consider a model that engenders the $Z_2\times Z_3$ symmetry, that is,
the discrete counterpart of $SU(2)\times SU(3)$. For this reason, we follow
the lines of Ref.~{\cite{bb00b} to introduce the model
\ben
\label{m}
{\cal L}&=&\frac{1}{2}\partial_\mu\sigma\partial^\mu\sigma
+\frac{1}{2}\partial_\mu\phi\partial^\mu\phi+\frac{1}{2}\partial_\mu\chi
\partial^\mu\chi
-V(\sigma,\phi,\chi)\nonumber\\
&+&i\bar{\psi}\gamma^\mu\partial_\mu\psi+m\bar{\psi}\psi
-f\sigma\bar{\psi}\psi
+\lambda(\phi+\chi)\bar{\psi}{\psi}
\een
This model contains three real scalar fields that couple among themselves via
the potential $V(\sigma,\phi,\chi)$, which is introduced below. Also,
there is a massive Dirac fermion $\psi$ that couples to the scalar
fields via the Yukawa couplings $f\sigma\bar{\psi}{\psi}$ and
$\lambda (\phi+\chi)\bar{\psi}{\psi}$ --- see Ref.{\cite{jack}} for
information on the behavior of fermions in the background of topological
defects generated by real scalar fields. 

The potential is chosen to provide the standard spherical soliton
star \cite{tdl1} with a network of domain walls on its surface. The sigma
field has to give rise to the host domain wall, which
should entrap the other two fields, which have to engender the $Z_3$ symmetry.
The model should be able to describe a spherical soliton star via the scalar
field $\sigma$, by breaking its $Z_2$ symmetry under the shift
$\sigma\!\to\!\sigma-(1/2)\sigma_0$, to entrap the other two fields $(\phi,\chi)$
with a $Z_3$ symmetry on its surface. We get to this model by considering
the potential 
\ben
\label{p}
V&=&\frac{1}{2}\mu^2\sigma^2(\sigma-\sigma_0)^2+\lambda^2(\phi^2+\chi^2)^2-
\lambda^2\phi(\phi^2-3\chi^2)
\nonumber\\
&+&\left[\lambda\mu(\sigma-\frac{1}{2}\sigma_0)^2-\frac{9}{4}\lambda^2\right]
(\phi^2+\chi^2)
\een
Here $\sigma\!=\!0$ and $\sigma\!=\!\sigma_0$ are the true and false vacua
corresponding to the standard soliton star.

The scenario for a soliton star to entrap a network of domain walls should
constrain the symmetries of the potential. According to the Euler theorem,
to tile the sphere with regular polygons of the same type we need (a) 4, or
(b) 8, or (c) 20 triangles, or (d) 6 squares, or yet (e) 12 pentagons.
These tiling show the cases where the polygons edges end in three-junctions
(a, d, e), four-junctions (b) and five-junctions (c). These configurations
are topologically equivalent to the five Platonic solids, respectively the
tetrahedron $\{3,3\}$, octahedron $\{3,4\}$, icosahedron 
$\{3,5\}$, cube $\{4,3\}$ and dodecahedron $\{5,3\}$; see
Refs. \cite{cox,fus} (here the notation $\{M,N\}$, stands for
regular $M$-gons and $N$-junctions). In this sense we can use
a potential with $Z_N$ symmetry between the fields $\phi$ and
$\chi$ (that must inhabit the interior of the domain wall formed
by the other field $\sigma$) to describe locally $N$-junctions on
the surface of a sphere \cite{bb00b} (the soliton star surface). We
have three possibilities of choosing $N$ ($N\!=\!3,4,5$), and we choose
$N\!=\!3$, that is, the $Z_3$ symmetry. While this is the minimal possibility
for the appearance of junctions, it is also the center of the $SU(3)$ group, 
that is the group of the strong interactions. Another possibility
with the symmetry breaking $Z_3\!\to\! Z_2$ was considered in \cite{bb00b}
in order to describe a tiling with 12 pentagons and 20 hexagons, that
resembles the fullerene structure of 60 carbon atoms.
The main mechanism for a domain wall to entrap a network
of domain walls was explored in Ref. \cite{bb00b}. The key point here is
that, as we shall see, on the surface of the soliton star
$\sigma\simeq(1/2)\sigma_0$, and at this place, the remaining
fields $(\phi,\chi)$ develop nonzero v.e.v. (condensate) with three 
different phases that contribute to form domain wall three-junctions
and then a network. We summarize this phenomenon as follows. 

We use the equations of motion to see that in the false and true vacua 
$\sigma\!=\!\sigma_0$ and  $\sigma\!=\!0$, respectively, the fields
$(\phi,\chi)$ turn out to be zero. For scalar fields $\phi$, $\chi\!=\!0$
and fermion field $\psi\!=\!0$ the theory (\ref{m})-(\ref{p}) allows the
field $\sigma$ to form a soliton solution. We note that such a solution
can be found by using the following first order differential equation
\ben
\label{edo}
\frac{d\sigma}{dR}=\mu\sigma(\sigma-\sigma_0)=W_\sigma
\een
where $W\!=\!\mu(\sigma^3/3-\sigma^2\sigma_0/2)$ can be seen as a
superpotential that define the potential $V(\sigma,0,0)\!=\!(1/2)W^2_\sigma$
\cite{bb}. The above Eq.~(\ref{edo}) can be integrated to give the solution 
\ben
\label{shell}
\sigma=\frac{\sigma_0}{2}\left[1-\tanh{\frac{\mu\sigma_0(R-R_0)}{2}}\right]
\een
This solution shows that at the surface $(R\simeq R_0)$ the 
$\sigma$ field goes to $(1/2)\sigma_0$. It represents approximately
\cite{kolb} a spherical wall (the soliton star surface) 
with surface tension \cite{bb}
\ben
\label{th}
t_h\simeq|W(\sigma_0)-W(0)|=\frac{1}{6}\mu\sigma_0^3.
\een

In the regime of $\sigma\simeq (1/2)\sigma_0$ the remaining scalar fields
$(\phi,\chi)$ engender $Z_3$ symmetry, and describe three-junctions of
domain walls which allow the formation of a network \cite{bbr00a}.
In the thin wall approximation each segment of the network can be
represented by a domain wall (kink) solution of the explicit form
\ben
\label{netw}
\phi&=&-\frac{3}{4}
\nonumber\\
\chi&=& \frac{3}{4}\sqrt{3}\tanh{\sqrt{\frac{27}{8}}\lambda (z-z_0)}
\een
The other segments are obtained by rotating the $(\phi,\chi)$ plane by
$120^o$ and $240^o$ degrees, respectively. Below we shall investigate
how the domain wall segments in the network may have normalizable fermionic
zero modes. The issue of whether or not we have normalizable zero modes on
the junction, (in \cite{ino} it was found nonnormalizable zero mode on BPS
junctions, in the supersymmetric context), does not affect our discussion
since the junction here is itself an approximately zero dimensional object,
and then with a negligible fermi gas if fermions may bind to it.

\subsection{Neutral Network on Fermion Soliton Stars}
\label{nn}

The spherical wall described by the $\sigma$ field is a non-topological
soliton. Thus it requires a conserved Noether charge of bosonic and/or
fermionic origin to stabilize it. According to the model used in this paper,
we are choosing fermionic charges to stabilize the soliton star.
The soliton star is then a fermionic soliton star.

In this case, in the false vacuum $\sigma_0$ the mass
of the fermion field goes effectively to zero if we assume
\ben
\label{fm1}
m-f\sigma_0=0 
\een

The soliton star becomes stable due to a three dimensional fermi gas
pressure (See, e.g., Ref. \cite{macp,bazm} for a similar discussion in
the context of {\it fermi balls}). This is the scenario of the standard
non-topological structure called fermion soliton star \cite{tdl2} (similar
studies for scalar soliton stars were considered in \cite{tdl3}), and now we
have to consider the effects of the other two fields $\phi$ and $\chi$ we
have already introduced in the model.

Regardless of the type of network that inhabit the soliton star, in general
the total energy of the system is $E\!=\!E_n+E_s+E_k$, which will be given below.
Since we are assuming that the fermions are in the interior of a spherical
false vacuum of radius $R$ in $(3,1)$ space-time dimensions, then the kinetic
energy of the confined fermions is
\ben
\label{e1}
E_k=\frac{Q}{R}, \qquad Q= \frac{1}{2}\left(\frac{3}{2}\right)^{5/2}
\pi^{1/3}N^{4/3}
\een
where $N$ is the fermion number. The surface of the soliton star 
contains the surface energy
\ben
\label{e2}
E_s=\alpha R^2, \qquad \alpha=4\pi t_h
\een
where $t_h$ is the surface tension of the soliton star. Finally, the energy
of the nested network is 
\ben
\label{e3}
E_n=\beta R, \qquad \beta=n\xi t_n
\een
where $n$ and $\xi R\!=\!d$ are the number and length of the segments in the
network, $t_n\!=\!(27/8)\sqrt{3/2}\lambda$ is the tension
of each segment \cite{bbr00a,bb00b}, and $\xi$ is a real constant.
The total energy is therefore
\ben
\label{e4}
E=\alpha R^2+\beta R+\frac{Q}{R}
\een
We minimize this energy (\ref{e4}) by using $\partial E/\partial R\!=\!0$,
which allows obtaining the critical radius  
\ben
\label{r}
R_0=\frac{1}{6\alpha}\left[A^{1/3}+\frac{\beta^2}{A^{1/3}}-\beta\right]
\een
where 
\ben
\label{A}
A=54Q\alpha^2-\beta^3+6\sqrt{3}\sqrt{Q(27Q\alpha^2-\beta^3)}\alpha
\een
We notice that in this scenario $Q$ turns out to have a lower 
bound, in order for a critical radius to exist. It is given by 
\ben
\label{Q}
Q\geq\left(\frac{\beta}{3}\right)^3\frac{1}{\alpha^2}
\een
which is non vanishing for $\beta\!\neq\!0$. In the limit of large $N$ 
the total energy of the configuration is
\ben
\label{EQ}
E\sim\alpha R_0^2=\alpha \left( \frac{Q}{2\alpha}\right)^{2/3}\sim N^{8/9}
\een
We note that the exponent of $N$ is lesser than unit. This means
that for large $N$ the energy of the soliton is always lesser than the
energy of the free particles, thus the stability of the soliton star
is ensured. This is the same limit obtained in \cite{tdl1,tdl2}. We conclude
that a neutral network does not contribute in the large $N$ limit.

Let us now consider the network contribution to the cold matter
upper mass limit $M_c$ in this scenario. We may estimate such a limit 
by simply equating the radius (\ref{r}) to Schwarzschild radius $R_s\!=\!2GM$.
First of all, we note that the minimum of the energy (\ref{e4}) at the
critical radius (\ref{r}) is the soliton mass
$M=3\alpha R_0^2+2\beta R_0$, where we have written the charge
$Q$ in terms of $R_0$, that is, $Q=(2R_0\alpha+\beta)R_0^2$.
Now we make $R_0\!\sim\!R_s$, which leads to
\ben
M_c&\sim&\frac{1}{12\alpha}\left(\frac{1+4\beta G}{G^2}\right)\nonumber\\
&\sim&(48\pi t_h G^2)^{-1}+[12\pi(t_h/t_n)G]^{-1}(n\xi)
\een 
where $n,\,\xi$ are numbers inherent to the type of network one is
considering. They contribute to higher {\it excitations} due to the network
at the surface of the soliton star. Notice that 
for $n\!=\!0$ (no network) $M_c$ reduces to the first term (``the fundamental
state") which is the same result found in Ref. \cite{tdl1}. Since for
$n\!\neq\!0$ the last term is positive we conclude that the network raises
the standard value of $M_c$ and then yields
heavier soliton stars. For a typical energy scale of the order of GeV and  
$\lambda\!=\!\mu\!=\!1$ we find the values for the tensions
$t_h\!=\!(1/6)(30\,{\rm GeV})^3$ and $t_n\!=
\!6\sqrt{3/2}(3/4)^2({\rm GeV})^2$ (here we have chosen 
$\sigma_0\!=\!30\,{\rm GeV}$). Now using the fact that 
$\sqrt{G}\!=\!l_p\!\simeq10^{-33}$ cm (the Planck length) and
$M_\odot\sim 10^{33}$ g we find
\ben
M_c&\sim&(30\,{\rm GeV})^{-3}l_p^{-4}+
(30\,{\rm GeV})^{-1}l_p^{-2}(n\xi)\nonumber\\
&\sim&(10^{15}+10^{-23})M_\odot
\een
Here we have dropped the factor $n\xi$ in the last step, since it can
change the mass of the soliton star at most for two orders of magnitude. 
This is because ${\sl max}(n)\!=\!30$ (i.e., the dodecahedron $\{5,3\}$ case)
and ${\sl max}(\xi)\!=\!2\pi$ (i.e., the largest arc on the sphere). This
result allows to conclude that in this model the network raises slightly
the mass of the standard soliton star. The radius size of this object
for $\sigma_0\!=\!30$GeV is $R_0\sim10^2$ light years.

\section{Fermion zero modes on the network}
\label{fnt}

In this section we discuss how to adjust the parameters in our model
in a way such that the fermions prefer to migrate from the false 
vacuum to the network. It is perfectly possible that 
the effective fermion mass inside the network which, in turn, is 
at the surface of the soliton star, goes to zero if we assume
the fine tuning
\ben
\label{fm2}
m-f\sigma_s-\lambda v_s=0 
\een
where $\sigma_s\!=\!(1/2)\sigma_0$ is the value of the $\sigma$ field
at the surface of the soliton star, and $v_s\!=\!3/4$ is the norm of
the vector field $(\phi,\chi)$ inside the network.

The way we couple fermions to the fields $(\phi,\chi)$ in the model
(\ref{m})-(\ref{p}) is standard, although it does not preserve the 
$Z_3$ symmetry that governs the $(\phi,\chi)$ portion of the model,
which allows the formation of a regular network of domain walls
as shown in Ref.~\cite{bbr00a}. With the alternative coupling
$\lambda\chi(\chi^2-3\phi^2)\bar{\psi}\psi$, the $Z_3$ symmetry should
be preserved, but in this case nonnormalizable zero modes like
$\psi(z)\!=\!\exp{(\pm C\, {\rm sech^2\, z})}\epsilon_{\pm}$, with
$m\!=\!(1/2)f\sigma_0$ ($C$ is a real constant.) would be present.
Similar conclusions were found in \cite{ino}. There, one has found
that an object preserving the $Z_3$ symmetry (BPS junctions, in global
supersymmetry) gets nonnormalizable zero modes as well.

The effective fermion mass given in terms of the background solution
(\ref{netw}) confined to the surface of the soliton
star can be written as
\ben
\label{FM3}
M_F(z)&=&m-\frac{1}{2}f\sigma_0\nonumber\\
&-&\left(\frac{3}{4}\right)\lambda\left[1
-\sqrt{3}\tanh{\sqrt{\frac{27}{8}}\lambda (z-z_0)}\right]
\een
In the above equation we see that $M_F(z_0)$ recovers
the left hand side of Eq.~(\ref{fm2}), allowing to conclude that
the effective fermion mass $M_F(z)$ goes to zero inside the network.
This means that the fermions prefer to live inside 
the network $(z\simeq z_0)$ rather than in the false vacuum
$\sigma\!=\!\sigma_0$. The fermionic zero mode inside the network
is described by
\ben
\label{zeromode}
\psi(z)&=&e^{ix^\nu p_\nu}\exp{\left(\pm\int_{0}^z{M_F(x)}dx\right)}
\epsilon_\pm\nonumber\\
&=&e^{ix^\nu p_\nu}
e^{\pm(m-f\sigma_s-\lambda v_s)z}\nonumber\\
&\times&\left[\cosh{\sqrt{\frac{27}{8}}\lambda(z-z_0)}\right]^{\pm\frac{1}
{\sqrt{2}}}\epsilon_\pm
\een
Here we have used $\gamma^\nu p_\nu\psi\!=\!0$ and $\gamma^z\epsilon\!=
\!\pm i\epsilon_\pm$
to solve the Dirac equation for the zero mode ($\nu\!=\!0,1$ is the
tangent frame. Also, $z$ is the coordinate transverse to each domain
wall at the surface of the soliton star and $\epsilon_\pm$ is a constant
2-spinor.) From Eq.~(\ref{fm2}), we see that the second exponential factor
in (\ref{zeromode}) does not contribute to the zero mode. Finally we find
that the only normalizable zero mode is
\ben
\psi(z)=e^{ix^\nu p_\nu}\left[\cosh{\sqrt{\frac{27}{8}}
\lambda(z-z_0)}\right]^{-\frac{1}
{\sqrt{2}}}\epsilon_\pm
\een
This shows that there are localized chiral fermion zero modes
\cite{mr3,jack,bazm,stoi} into each domain wall segment of the network.
In other words, there are localized massless fermions only on the network.

\subsection{Charged Network on Neutral Soliton Stars}
\label{cn}

Now we are ready to present another scenario. Let us suppose that all
the fermions feel an attractive strong force so that they are forced
somehow to migrate from the false vacuum to the nested 
network on the surface of the soliton star. This is exactly what is
experienced by the fermion zero modes that we have just treated. In this
case we should replace the former tridimensional fermi gas by another
Fermi gas, approximately one-dimensional, which spreads along the network
conserving the fermion number $N$. Thus, all we have to do in the
investigation done in the former Sec.~{\ref{nn}} is to replace
Eq.~(\ref{e4})
by
\ben
\label{e4.1}
E=\alpha R^2+\beta R+\frac{\gamma}{R} ,
\qquad \gamma=\frac{\pi N^2}{4\xi}
\een
where we have used the expression for one-dimensional fermi gas \cite{mr3}
\ben
\label{e4.2}
E=\frac{\pi N^2}{4L}
\een
where $L\!=d\!=\!\xi R$ is the length of the segments in the network.
In the large $N$ limit we can also use (\ref{EQ}) to obtain
\ben
\label{EG}
E\sim\alpha R_0^2=\alpha
\left( \frac{\gamma}{2\alpha}\right)^{2/3}\sim N^{4/3}
\een
where we have set $Q\to\gamma$. Now, since the exponent of $N$ is greater
than unit, the energy of the soliton for large $N$ 
is always larger than the energy of the free particles and then the
stability of the soliton star is not ensured anymore.  

\section{Conclusions}
\label{conc}

In this paper we have presented a model which can describe two totally
different scenarios, described in Sec.~\ref{nn} and in Sec.~\ref{cn}.
In the first scenario, although the neutral network
slightly increases the cold matter upper mass limit of the standard soliton
star, it imposes a new lower bound on the charge of the fermionic soliton
star. In the second scenario, the entrapment of the Fermi gas inside the
network changes significantly the behavior of the nontopological soliton,
destabilizing the soliton star.
In the equilibrium stage of the fermion migration, the false vacuum gets
neutral and the network gets charged with massless fermions. This result
lead us to the conclusion that in the scenario of Sec.~\ref{cn}, the
formation of a charged network at the surface of the soliton star can
destabilize it. This is because the localized fermion zero modes on the
network make it energetically favorable to the soliton star to decay into
free particles. They can decay fast before the cold matter upper mass limit
$M_c$ is achieved. These two scenarios can be thought of as two possible
different experiences that a soliton star can suffer in the cosmological
evolution. Suppose that the soliton star and the network appear at distinct
critical temperatures $T_s$ and $T_n$, respectively (see Ref.~{\cite{bb97}},
for a study about critical temperature and defects formation.)
Thus for $T_s$ sufficiently
larger than $T_n$ the soliton star can start its formation process by
collapsing, approaching to the Schwarzschild radius before the strong
force due to the fermionic zero mode of the network acts. If the network
appears later, it will be ``frozen'' together with the other constituents
of the soliton star due to a strong gravitational force, leading to
the scenario of Sec.~\ref{nn}. On the other hand, if $T_s\!\sim\!T_n$ one
may induce the scenario described in Sec.~\ref{cn}. In this scenario,
if we consider that the soliton star may form cosmic objects, we see
that the more the soliton star has the experience \ref{cn} the less
these cosmic objects are formed in the universe.

The new objects that spring in the present work may be of
particular interest to astrophysical and cosmological applications,
as for instance in the recent investigations introduced in
Refs.~{\cite{dtcl,bb00b,fus}}. Furthermore, these objects provide alternative
routes to the Fermi balls examined in Refs.~{\cite{macp,bazm}}, and as
such they may furnish distinct view of the scenario there discussed.
Another line of investigation may follow Ref.~{\cite{pet96}}, where one
examines the behavior of surface current-carrying domain walls, the current
being of bosonic origin, which appears in a model engendering the
$U(1)\times Z_2\to U(1)$ symmetry, with $U(1)$ global. The model of
Ref.~{\cite{pet96}} is inspired by the case of cosmic strings introduced in
Ref.{\cite{wit85}}. Our model provides another possibility, where fermionic
current may flow on the surface of a wall that hosts a network of domain
walls, so that we could ask how the presence of the nested network
would contribute to change the fermionic behavior in the wall. Some of
these issues are presently under consideration, and we hope to report
on them in the near future.

\acknowledgments

This work is supported in part by DOE
grant DE-FG02-95ER40893 and NATO grant 976951. FAB would like to thank
Conselho Nacional de Desenvolvimento Cient\'\i fico e
Tecnol\'ogico, CNPq, Brazil, for support, and Department of Physics and
Astronomy, University of Pennsylvania, for hospitality. DB would like
to thank CNPq and PRONEX for partial support.
 
 
\end{document}